# The Learning Technique of the SageMathCloud Use for Students Collaboration Support

Maya Popel[1], Svitlana Shokalyuk[2], Mariya Shyshkina[1]

[1]Institute of Information Technologies and Learning Tools of NAES of Ukraine,
9 M.Berlynskoho St., Kyiv, Ukraine

`{popel, shyshkina}@iitlt.gov.ua`
[2]Kryvyi Rih State Pedagogical University
54 Gagarina Ave., Kryvyi Rih, Ukraine

`ksv_ipm@mail.ru`

**Abstract.** The article describes the advisable ways of the cloud-based systems use to support students' collaboration in the process of math disciplines learning. The SageMathCloud-based component that aggregates electronic resources for several math disciplines training is introduced. The learning technique of the SageMathCloud use in the process of educational staff training is proposed. The expediency of this technique implementation for more active take up of innovative approaches, forms and methods of math training with the use of the cloud-based tools is substantiated. The experimental results of the SageMathCloud learning component introduction research along with the methods of its use that were elaborated in the course of the study are presented. The use of the evidence-based technique as improving the educational environment of the university, empowering access to electronic learning resources in the course of math training and engaging with this the educational community and also rising their ICT competence is grounded.
**Keywords:** cloud computing, ICT-based tools, learning environment, teamwork, learning collaboration.
**Key Terms:** Methodology, InformationCommunicationTechnology, ICTTool

## 1 Introduction

### 1.1 The Problem Statement

Application of the cloud computing (CC) technology in education is to provide the comprehensive development of learner's personality and competencies according to the individual needs and requirements of a society. Its implementation is associated with the objective need to create an open learning environment where students become the active subjects of purposeful search and creation of knowledge. In this concern, the research and implementation of essentially new forms of learning and educational activity are necessary. The learner-centered approach and adaptive learning technologies empowering students and teachers collaboration and increasing the share of knowledge become the central point.

With the use of the cloud technology the better results of learning may be achieved, providing each student with his own educational trajectory using adequate technological solutions. There are new opportunities for differentiation and individualization of the learning process, its flexible adaptation to the needs and individual characteristics of the students [2].

The cloud technology emergence alters our understanding of the hardware and software use and also the data storage and access in the learning process.

Really, the use of the cloud services enables the individuals to acquire and process their data, perform the calculations, make certain adjustments referring to them via the Internet. The user has no need to worry about software installing and updating, memory limits, the use of data-storage devices and processing.

And also there is no need to use super-power devices for training. It's enough to have access to regular notebook, Smartphone or any other device linked to the Internet, having free space to store data. The learning content and tools may be delivered at any time to any place due to the user demands. There is a problem to bring cloud computing technologies to educational practice considering the better ways of learning tools selection and pedagogical personnel training. There is a need to develop the learning techniques oriented to the most powerful and open accessed tools.

In this regard, the prospects of the SageMathCloud service implementation being quite powerful and yet free accessed is an actual problem of theory and methods of using ICT in teaching mathematical disciplines.

### 1.2 The State of the Art

In the recent years the wider use of the cloud-based tools and services became the central point, forming an ICT platform of modern educational systems and learning environment [1; 5; 13; 15; 16].

For example, M. A. Kislova considers the mobile learning environment for mathematics learning in the university that includes the tools for various mathematical activities support, mobile learning communication, enhancing the interaction between students and a teacher [4].

The several cloud-based tools are considered by M. A. Kislova and K. I. Slovak [3] who offered to use them in combination with other learning tools. Among them there are such as: Google Apps for Education, Office 365, ThinkFree Online. The benefits of their use for teaching mathematical disciplines are outlined.

In the current research of Systems of Computer Mathematics (SCM) the recent tendency was determined that the popular mathematical software tools appear at cloud versions, for example such as Maple Net, MATLAB web-server, WebMathematica and others. The main differences between SCM and web-based SCM and also the cloud-based SCM are highlighted in more detail in [9]. The SageMathCloud is an example of such kind of software as the cloud-based version of Web-SCM Sage [9].

Some experience of using the cloud services in educational in Volodymyr Gnatiuk National Pedagogical University of Ternopil is reported in [6]. The cloud-based infra-structure with the Google Apps cloud services integrated into the learning environment of physics and mathematics faculty were described [6].

The most studies of Ukrainian researchers on the matter of research in the resent years were focused on the principles, approaches and models design of the environments of higher education, which included the cloud services [9; 10; 12].

The experience of using the SageMathCloud in teaching the optional course in Tutek Željka Croatia's University of Zagreb in the period of 2015-2016 years presented in [14] shows that the primarily attention is necessary to focus on:

− acquiring the skills of using the mathematical software for numerical and symbolic computation;
− supporting mathematical concepts teaching (including courses in Analytic Geometry, Linear Algebra and Calculus).

Usually the cloud services can be used for data visualization and calculation, particularly for solving problems within the certain discipline and organization of individual and collective work, the student learning monitoring.

So the advisable ways of the SageMathCloud use in pre-service math teachers training need more careful research. The important topics are the following:

1. Organization of educational communication;
2. Keeping individual and group forms of learning activities (class and extracurricular);
3. Keeping a learning management;
4. Providing visibility by constructing different interpretations of mathematical models, visualizing mathematical abstractions etc.;
5. Ensuring availability and research opportunity by using a common interface of learning environment objects access and reliable software with open source;
6. Increasing the temporal and spatial mobility;
7. Forming a unified learning environment with content developed in learning.

### 1.3　The Purpose of the Article

The *main purpose* of the paper is the analysis of the most appropriate ways of the SageMathCloud learning component use to support collaborative work in the process of mathematical disciplines learning and justification of the technique of its application for training lecturers and academic staff in the pedagogical university.

## 2　Presenting the Main Material

The use of ICT affects the content, methods and organizational forms of teaching and managing educational and cognitive activity that requires new approaches to educational process arrangement. Therefore, the formation of modern cloud-based systems for supporting teaching and research activities and students' collaboration on the learning projects should be based on appropriate innovative models and method-ology that can ensure a harmonious combination and embedding of various networking tools into the information-educational environment of higher education. The model of the SageMathCloud use to support various forms of collaboration in the process of training of pre-service teachers of mathematics (Fig. 1) consists of the following components: the organizational, the content and the technological ones. The model includes teachers' and students' individual or collective spaces consisting of some structural elements. The combination of certain elements provides different types of interaction organization (between students and a teacher and between students with each other) among them such as individual and group work of students, students and teachers cooperation and active communication.

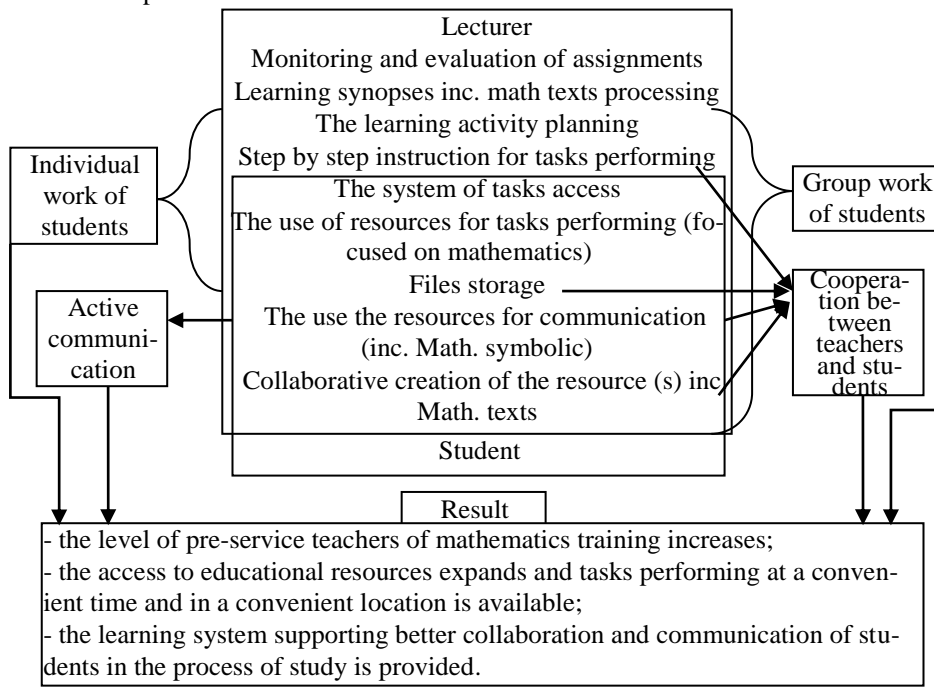

**Fig. 1.** The model of the SageMathCloud use to support various forms of collaboration in the process of training of pre-service mathematics teachers

In this connection, in 2014-2016 in the Institute of Information Technologies and Learning Tools of NAES of Ukraine the research study devoted to the use of the SageMathCloud in pre-service teachers of mathematics training was carried out (M.Popel) [11]. The main idea of the study was the provisions that purposeful, scientifically grounded system of training using the cloud-based environment tools and technologies integrated into the process of educational staff training at the university is the precondition for improving the research process, widening access to electronic resources and services, rising the ICT competence and learning outcomes of students and teachers.

In the course of research the cloud-based learning component of the information and educational environment included the SageMathCloud was elaborated and meth-ods of its use were designed. It was introduced through the workshops, webinars, seminars and other scientific and methodical and learning events conducted by the Institute [11].

### 2.1 The Technique of the SageMathCloud Use

**The Target**

The aim of the SageMathCloud learning component use is to provide the organizational and technological support for students' collaboration in the process of study, expanding access to qualitative electronic educational resources (EER), increasing professional and ICT competencies.

The target group: educational and pedagogical and research personal.

**The Learning Content**

The learning content to be provided so as to incorporate the learning technique of the SageMathCloud use occupies the topics aimed at formation of ICT competence of research and educational staff, ICT departments employees and also students on the use of the cloud-based systems and services in the research and educational process. Some learning modules of such kind may be the elements of training, re-training of academic and lecturers staff content.

**The Technological Aspect of Use**

The advisory teaching methods are: explanatory learning; partly research learning; problem learning; research learning.

The organizational forms of learning are: lectures; individual, practical, laboratory work; training sessions; seminars, webinars, web conferences; individual consultations.

Among the innovative forms of learning that can be realized only in the cloud-based environment the combined training is appropriate. This kind of learning integrates internal and remote forms of work. In the process of training the situational e-learning network is designed and its members are motivated to maintain the collective activity by a common scenario. The organizer gives the examples of successful training activities (as the previous material templates of training tasks and also interactively as using the visual and auditory tools for submitting worksheets and tasks monitoring and organizing ICT-mediated learning management). The activity increase occurs due to the involvement of the training experts.

The main important learning tools elaborated with the use of the SageMathCloud are:

— The working sheets on which students perform their activity with the construction and study of mathematical models;
— The chat room used for discussions of the simulation process and results;
— The tools to support learning activities (resources such as course, tasks);
— The tools for creating mathematical texts (tex) and hypertext (html);
— The mobile tools for other mathematical activities access.

**The Ways of Implementation**

The implementation of the proposed method in the learning process may occur in two ways:

1. 1. Through the introduction of the special learning course "Cloud Computing Technologies in Educational Activity", as an element of the content of training, retraining of academic and lecturers staff [5].
2. Through the introduction of a system of training (see Table 1) such as seminars, webinars, individual consultations that can be carried out during the pilot experimental study (project) on the deployment of a cloud-based information and educational environment at the university.

**Table 1.** Plan of trainings

| № | Topic Title | Number of hours |
|---|---|---|
| 1. | Introduction. The advisable ways to use the SageMathCloud in learning mathematical disciplines | 2 |
| 2. | Organization of collaboration in the SageMathCloud | 2 |
| 3. | Construction of lecture demonstrations | 2 |
| 4. | Creating of dynamic models and animations | 2 |
| 5 | Examples of SageMathCloud use in teaching of certain math disciplines | 2 |
| | Total: | 10 |

The use of the SageMathCloud in teaching mathematical disciplines should be primarily targeted to:

- The acquisition of skills of individual use of the mathematical software for numerical and symbolic computation;
- Supporting of mathematical concepts learning (including courses in Analytic Geometry, Linear Algebra and Calculus).

**The Tools of Collaboration Support in the SageMathCloud**

The main activity type in the SageMathCloud is going on with the project. The user can create any number of independent projects or personal workspaces, to save resources of various types, including:

- SageWorksheet (*.sagews);
- LaTeX Document (*.tex);
- Manage a Course (*.course);
- Chatroom (*.sage-chat);
- Folder; etc.

Collaboration with the use of the SageMathCloud-project resources can be organized in of two possible ways:

- at the level of a single resource, including worksheets;

– at the level of a project as a whole.

Sharing access at the level of a single resource is nothing more than a webpromulgation of resource content. Moreover resource is made public for all Internet users who have a link to it and they can operate with it in the mode "read only". The main disadvantage of this publication is that on the one hand a user as a "reader" is not able to manage calculations on the worksheet (even if it contains standard elements of management) on the other hand this worksheet can be copied or downloaded along with its full source code that does not guarantee security and protection of user copyrights.

Organization of collaboration at the level of the whole project is possible the other way using a courses manager (Fig. 2), or without it.

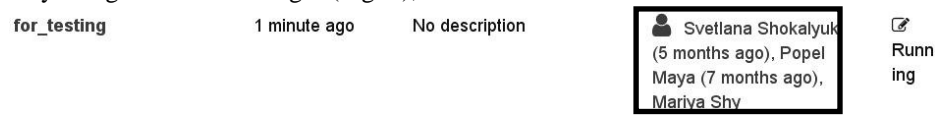

**Fig. 2.** Overall view of the joint project

In case of collaboration without the course manager all participants have the opportunity to:

– collaborate using existing (offered by a teacher) educational resources of a project;
– add new training resources;
– invite other members to collaborate on the project (Fig. 3);
– communicate to each other and a lecturer by means of text chat or video chat.

While working with a learning resource in real time the users are able to see cursor positions of each other and the icons of those project participants who addressed to a file at this moment, as like as it happens if working collaboratively with a Google document.

The contribution of each project participant in the tasks can be viewed at the project backup (Fig. 3) or in the pages of history of the project or a worksheet.

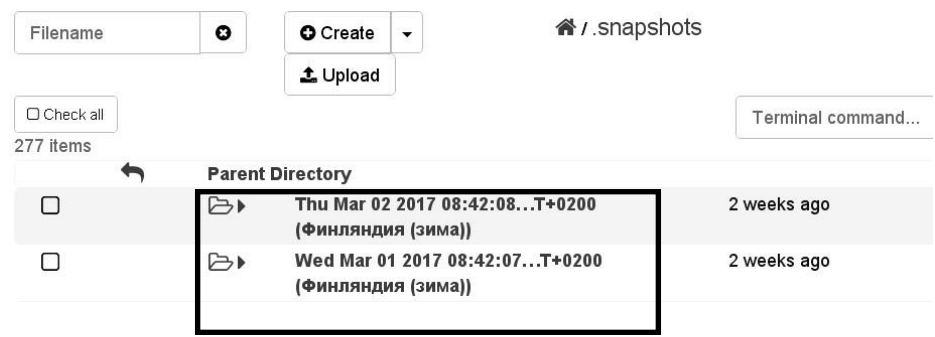

**Fig. 3.** Page of the backup versions of the «Backups» project

Backups of a project («Backups») are saved every 5 minutes and can not be completely removed by the user.

Contribution of each member of the joint project into the tasks solving may be reviewed on the history pages of the project «Log» (Fig. 4): creating files, folders, deleting, moving, copying, opening, renaming and more.

To review the whole history of changes of the worksheet content the page that opens after clicking «TimeTravel» may be used (Fig. 5). In the history file the records of all changes made by the author of a project or by any other participant are saved. Moving the slider, you can view all the changes that were made on the worksheet.

Initial position of the slider corresponds to the file creation.

**Fig. 4.** Page of project history «Log»

If the file was not changed nor edited, it is indicated «Revision 0". For each change the date and time are shown. The last position of a slider reflects the recent changes that were made when editing the file.

**Fig. 5.** The page of worksheet history «Time Travel»

Communication between the project participants in real time is possible by means of traditional "rooms" of text chat (Fig. 6), video chat or on the pages of the sage-chat resource type. Text messages can be formatted with HTML tags and commands of Markdown. The message with mathematical content may be performed in usual mathematical notation using commands of LaTeX. In addition, for a resource type sage-chat the function of history viewing is available.

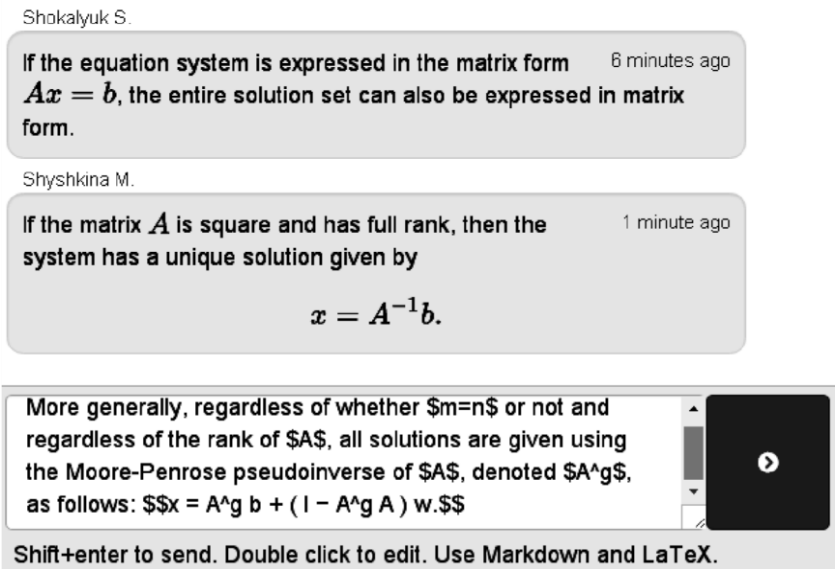

**Fig. 6.** The traditional "room" of a text chat

For the organization of individual collaboration with each student, the teacher creates a separate project and fills it with educational resources and gives access only to a specific user (the student). As a result, the number of projects created by the students is determined by the number of academic groups. In this case the teacher is forced to spend a lot of time to do the same actions dozens of times (as it is not possible to copy the whole SageMathCloud). The process may be automated by using the special SageMathCloud-resource namely the course Manager.

This method of training with the use of the SageMathCloud provides opportunities for teachers to control the process of tasks performing by the students and also the evaluation. The page of the training course Manager is shown in Fig. 7.

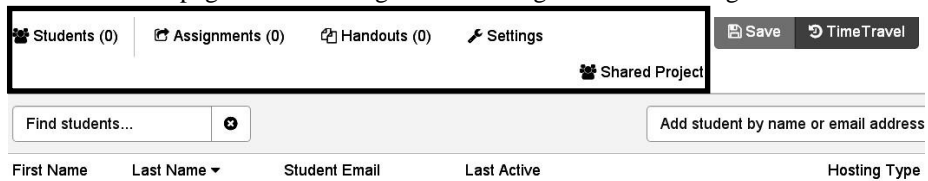

**Fig. 7.** The page of the training course manager

There are the next control elements on this page:

— the page label «Students» (for management of students actions);
— the page label «Assignments» (for management of educational activities within the course);
— the page label «Handouts» (for management of educational materials that are in folders);

— the page label «Settings» (to adjust the settings of the training course);
— the page label «Shared Project» (to create a project common for all students).

A typical sequence of actions that envisages the teachers work using training courses manager is as follows:

1. Inclusion of each student to the course (footprint refer to the page «Students»). When a user is included into the course as a student, a personal project is automatically created, with access to be opened to teachers and students.
2. Filling the training course with materials (hereinafter - management) is at «Assignments».
3. Copy the educational materials for all students to their projects is made simultaneously by clicking the button «Assign ...».
4. When copying educational materials at least to one student above the list of students in the box of general settings of folders the buttons appear that allow to assign tasks to all students of the course simultaneously, collect the data about tasks performing ( «Collect») and make the resulting assessment report ( «Enter grade»).
5. Educational materials that have been copied, you can update as needed.
6. On the page «Settings» of the general settings of a resource a teacher may be interested in the following blocks:

— «Export grades» - clicking on the button «CSV file ...» or «Python file ...», in the project structure will be created a file with indication of assessments of all students;
— «Customize email invitation» makes the text message concerning inclusion of a user into the course, which will go to the user mailbox, if he is not registered in the system (the text message can be changed by clicking on the button «Edit»).

On the page «SharedProject» a shared project can be created for all students of the training course, in which each student automatically becomes a member.

**Organization, Conduct and Results of Experimental Work**

The aim of pedagogical experiment was to justify empirically the theoretical assumption that the proposed pedagogical approach was really effective.

At the ascertaining stage of experimental research the analysis of existing equipment and logistics was made for it to be sufficient for the research; the current state of teaching staff readiness to use the cloud services in their professional work was estimated, appropriate tools were chosen. It was presupposed that the mastery of pedagogical staff in using the cloud services to support mathematical activity, including the SageMathCloud service, was a prerequisite for improving the quality of learning results.

At the ascertaining stage of the experiment some problems of specific professional competencies formation, in particular ICT competence were found.

Teachers would liked to improve their knowledge through greater use of cloud-based ICT in their professional activities.

Based on the data presented in Table 2, we may check the reliability of the hypothesis of no statistically significant differences between the levels of formation of ICT competencies of teaching staff in experimental and control groups. The chi-squared test was used. As the critical value of chi-square as 5.99 was chosen.

**Table 2.** Average percentage levels of ICT competence in the control groups (CG) and experimental groups (EG), the results of initial and final measurement

| Equal | Percentages | | | |
|---|---|---|---|---|
| | CG | | EG | |
| | primary measurement | final measurement | primary measurement | final measurement |
| High | 10 | 5 | 11 | 15 |
| Average | 41 | 42 | 51 | 63 |
| Low | 49 | 56 | 38 | 22 |

Defining empirical value of 2.53, we can see that it is in the area of insignificance and the hypothesis that the proportion of teaching staff with the appropriate level of ICT competence is the same in experimental and control groups by the results of entrance control is accepted.

After checking the authenticity of the hypothesis of statistically significant differences between the levels of ICT competencies of teaching staff in control and experimental groups at the end of the experiment the following results were obtained: 23.98> 5.99, where 23.98 is the empirical significance for hypothesis testing. That hypothesis is accepted that the proportion of teaching staff of the experimental group with high and medium level of ICT competence is higher than that in the control group.

Comparing the levels of ICT competence in the early formative stage and at the end of the experiment (Fig. 8) shows the increase of the proportion of teaching staff with the high and medium levels of ICT competence.

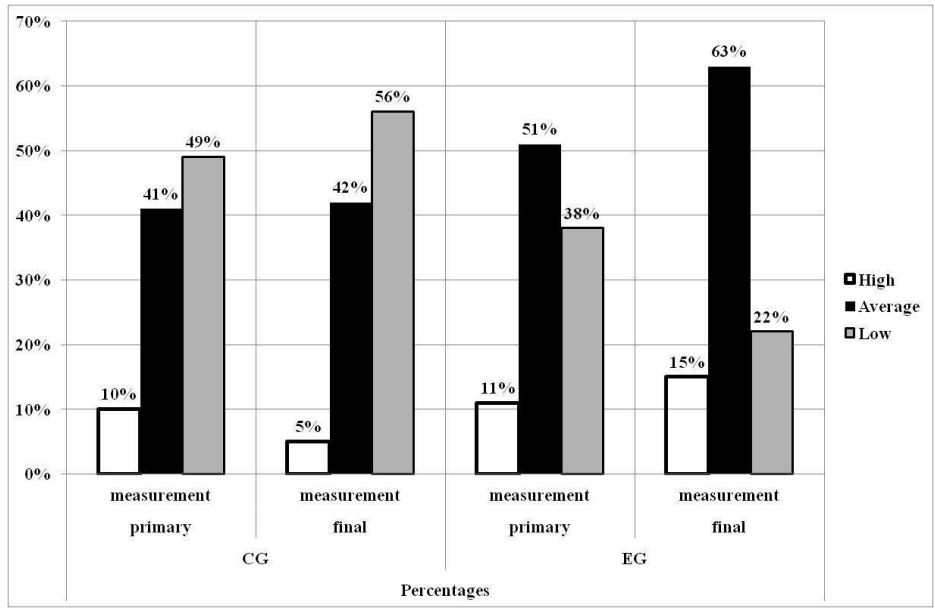

**Fig. 8.** Percentage distribution of ICT competency levels in the control and experimental groups, the results of initial and final measurement

Analysis of the formative stage of pedagogical experiment showed that the distribution of the levels of ICT competences in the experimental and control groups of teaching staff have statistically significant differences due to the introduction of the developed technique, which confirms the hypothesis of the study.

## 3      Conclusions

The introduction of the cloud-based tools and components into the learning process contributes to the improving of the educational environment of the university, increasing access to electronic resources and services, wider use of network technologies and tools in pedagogical practice. This is possible due to the application of experimentally grounded methods of the design and use of the cloud-based learning components on the basis of the SageMathCloud to support students collaboration in the process of training.

## References

1. Cusumano, M.: Cloud computing and SaaS as new computing platforms. In: Communications of the ACM, 53.4, pp. 27-29. (2010)
2. Hrybyuk, O.O., Demyanenko, V.M., Zhaldak, M.I., Zaporozhchenko, Y.H., Koval, T.I., Kravtsov, H.M., Lavrentyeva, H.P.,Lapinskyy, V.V., Lytvynova, S.H., Pirko, M.V., Popel, M.V., Skrypka, K.I., Spivakovskyy, O.V., Sukhikh, A.S., Tataurov, V.P., Shyshkina, M.P.:

Zhaldak, M.I. (eds.) The system of psychological and pedagogical requirements to the means of ICT for educational purposes. Kyiv (2014) (in Ukrainian)

3. Kislova, M.A., Slovak, K.I.: Cloud means of teaching mathematical disciplines. In: New computer technology, vol. XI, pp. 53-58. SHEE "National University Krivyi Rig", Krivyi Rig. (2013) (in Ukrainian)
4. Kislova, M.A.: Development of a mobile learning environment in higher mathematics in training of Electromechanics Engineers. P. 21. (2015) (in Ukrainian)
5. Maschietto, M., Luc Trouche: Mathematics learning and tools from theoretical, historical and practical points of view: the productive notion of mathematics laboratories, ZDM 42.1, pp. 33-47. (2010)
6. Oleksyuk, V.: Experience of the Integration Cloud Services Google Apps into Information and Educational Space of Higher Educational Institution. In: J. Information Technologies and Learning Tools, vol. 35, № 3, http://journal.iitta.gov.ua/index.php/itlt/article/view/824/63 (2013) (in Ukrainian)
7. Popel, M.: The Methodical Aspects of the Algebra and the Mathematical Analysis Study Using the Sagemath Cloud. In: Informational Technologies in Education, 19, pp. 93-100. (2014) (in Ukrainian)
8. Shokalyuk S.: Zhaldak, M.I. (eds.) Fundamentals SAGE. Kyiv (2008) (in Ukrainian)
9. Shyshkina, M.P., Kogut, U.P., Popel, M.V.: Systems of computer mathematics in the cloud-based learning environment of the educational institution. In: Science and Education a New Dimension. Pedagogy and Psychology. 27 (II(14)), pp. 75-78. (2014), http://lib.iitta.gov.ua/6499/1/article-science-edu.pdf (in Ukrainian)
10. Shyshkina, M.P., Popel M.V.: The cloud-based learning environment of educational institutions: the current state and research prospects. In: Information Technologies and Learning Tools. 5(37). (2013), http://journal.iitta.gov.ua/index.php/itlt/article/view/903/676 (in Ukrainian)
11. Shyshkina, M.P.: Formation and development of cloud-based educational and scientific environment of higher education. UkrINTEI, Kyiv (2015) (in Ukrainian)
12. Spirin, O.M., Demyanenko, V.M., Shyshkina, M.P., Zaporozhchenko, Y.H., Demyanenko, V.B.: Models harmonization of network tools of information technology maintenance processes of educational and cognitive activity. In: Information Technologies and Learning Tools., 6 (32). (2012), http://journal.iitta.gov.ua/index.php/ itlt/issue/archive (in Ukrainian)
13. Turner, M., Budgen, D., Brereton, P.: Turning software into a service. In; Computer, 36 (10), pp. 38-44. (2003)
14. Tutek, Ž.: Nastava matematike na SageMathCloud platform. In: Biljanović, P. (eds.) MIPRO 2016 Proceedings, pp. 1215-1217. Hrvatska udruga za informacijsku i komunikacijsku tehnologiju, elektroniku i mikroelektroniku, Rijeka (2016)
15. Vaquero, L.M.: EduCloud: PaaS versus IaaS cloud usage for an advanced computer science course. In: Education, 54.4, pp. 590-598. (2011)
16. Wick, D.: Free and open-source software applications for mathematics and education. In: Proceedings of the twenty-first annual international conference on technology in collegiate mathematics, pp. 300-304. (2009).